\documentclass{article}
\usepackage{amsmath}
\usepackage{mathtools}

\newlength{\defbaselineskip}
\newcommand{\setlinespacing}[1]%
           {\setlength{\baselineskip}{#1 \defbaselineskip}}

\usepackage[all]{xy}
\usepackage{amsmath,amssymb,amsfonts,amsthm,mathrsfs,dsfont,mathtools}
\usepackage[english]{babel}
\usepackage{graphicx,stackrel}
\usepackage{multirow,rotating,paralist}
\usepackage[toc,page,header]{appendix}
\usepackage[all]{xypic}
\usepackage{graphics}
\DeclareMathAlphabet{\mathpzc}{OT1}{pzc}{m}{it}
\usepackage{minitoc}
\usepackage{extarrows}
\hoffset - 1.5cm
\topmargin - 1.5cm
\newcommand\undersym[2]{\raisebox{-3pt}{\tiny$#2$}{\kern-5pt}\mbox{$#1$}}
\newcommand{\edf}{\ {\mathop{=}\limits^{^{def.}}}\ }
\title{\textbf{Torsion and Problems of Standard Cosmology}}
\author{M.I.Wanas$^{1,2,3}$ and H.A.Hassan$^{1,2}$}
\date{ }
\begin{document}
\maketitle
\begin{center}
    $^1$ Astronomy Department, Faculty of Science, Cairo University, Giza, Egypt, Email:wanas@frcu.eun.eg \end{center}
   \begin{center} $^2$ Centre for Theoretical Physics, The British University in Egypt, Cairo, Suez Desert Road, El Sherouk City, Email:mamdouh.wanas@bue.edu.eg;  hassanahh@gmail.com \end{center}
   \begin{center} $^3$ Egyptian Relativity Group (EGR), URL: http://www.erg.eg.net/index1.htm
\end{center}
\large{
\begin{abstract}
\hspace{12pt}
\setlinespacing{2.5}
\textbf{The field equation of orthodox general relativity are written in the context of a geometry with non-vanishing torsion, the Absolute Parallelism (AP) geometry. An AP-structure, with homogeneity and isotropy, is used for cosmological applications. The resulting dynamical equations are those of FRW-Standard cosmology, which have many problems e.g. singularity, particle horizons, ...etc. We suggest a new scheme for investigating the effect of torsion on the dynamics of FRW-Cosmology, without changing the basic structure of general relativity. It is shown that some of these problems will disappear if the torsion, associated with AP-structure used, is inserted in to the dynamical equations. Diagnose shows that problems arise when GR is written in the context of a geometry with vanishing torsion, the Riemannian geometry. This reflects the importance of using more wider geometries in studying physical phenomena. }
\end{abstract}
\setlength{\lineskip}{5pt plus 12pt minus 0pt}
\lineskiplimit=\baselineskip
\section{Introduction}
\setlinespacing{2.5}
\hspace{12pt}
 Standard cosmology is a branch of science that deals with the structure and evaluation of the Universe as one system. Theoretically, it depends mainly on the General theory of Relativity (GR). Just after its birth in 1915, GR has succeed in predicting important features of our Universe (e.g. The expansion of the Universe, the Cosmic Microwave Background Radiation(CMBR), the abundance of light elements,...etc). Some of these predictions have been directly confirmed by observations afterwards \cite{Hubble1929, Alpher21948, Alpher1948, Penzias1965}. However, standard cosmology suffers from some problems as singularity, particle horizons, flatness, and the accelerating expansions of the Universe \cite{Riess1998, Perlmutter1998},...etc. Many authors have attempted to get rid of these problems by reinserting the cosmological term in GR (cf. \cite{Arbab2003}), writing alternative theories depending on the curvature scalar R, f(R) theories (cf. \cite{Sotiriou2010, Faraoni2008, Thomas2009}), or inventing some scenarios as inflation  \cite{Guth1981, Linde1982}, baryogenesis \cite{Trodden2003}. All the above mentioned attempts have been done in the context of Riemannian geometry.
 \paragraph{}
Another class of attempts has been done by writing other theories, alternative to GR, in the context of geometries with non-vanishing torsion (T). This class is known in the literature as f(T) theories (cf. \cite{Myrzakulov2011, Wu2011, Gamal2012, Dent2011, Poplawski2010}). And also dealing with f(R) and f(T) theories together in one theory called F(R,T) gravity (\cite{Myrzakulov:2012}).

\paragraph{}
The aim of the present article is to study the effect of torsion on the dynamics of the Friedmann-Roberttson-Walker(FRW)-Cosmology. For this aim, the article is arranged as follows. In Section 2 we overview briefly the main feature of a geometry with non-vanishing torsion, the Absolute Parallelism (AP-)geometry. In the $2^{nd}$ part of this Section we use an AP-structure, satisfying the cosmological principle and evaluate tensors necessary for the present application. In the last part of this Section we write the field equations of GR in the context of the AP-geometry. In Section 3, we investigate the effect of presence of a non-vanishing torsion on the FRW-dynamics. Discussion and some concluding remarks are given in Section 4.
\section{ A Geometry with Non-vanishing Torsion}
\hspace{12pt}
In the present Section we are going to use a simple type of geometry with non-vanishing torsion, the AP-geometry (cf.\cite{Wanas2002, Wanas2000a, Youssef2006}). We are going to review briefly some of its properties and the geometric entities necessary for the present work. Calculations in the AP-geometry are easier compared to other types of geometries with non-vanishing torsion.
\subsection{The AP-Space}
\hspace{12pt}
In 4-dimensions, the structure of an AP-Space is defined completely in terms of a tetrad vector field $\undersym{\lambda}{i}^{\mu}$(the building blocks of any AP-structure), where $i (=0,1,2,3)$ represents the vector numbers and $\mu (=0,1,2,3)$ denotes the coordinate components of the vector. It is assumed that the tetrad vectors are totaly independent i.e. the determinant $\lambda(\edf \| \undersym{\lambda}{i}^{\mu} \|)$  is non-degenerate. This implies that, for any tetrad vector field there is a conjugate $\undersym{\lambda}{i}_{\nu}$ such that \footnote[1]{In the present article we are going to use Latin (mesh) indices for the vector numbers and Greek (world) indices for coordinate components. Einstein summation convention is carried out over Greek indices in the usual manner, while for Latin indices, it is carried out over repeated indices wherever their positions.}
\begin{equation}
\undersym{\lambda}{i}^{\mu} ~ \undersym{\lambda}{i}_{\nu} = \delta^{\mu}_{\nu},
\end{equation}
\begin{equation}
\undersym{\lambda}{i}^{\nu} ~ \undersym{\lambda}{j}_{\nu} = \delta_{ij}.
\end{equation}
Using the tetrad vector field and its conjugate we can define the following $2^{nd}$ order symmetric tensors,
\begin{equation}\label{q3}
g^{\mu\nu} \edf \undersym{\lambda}{i}^{\mu} ~ \undersym{\lambda}{i}^{\nu},
\end{equation}
\begin{equation}\label{q4}
g_{\alpha\beta} \edf \undersym{\lambda}{i}_{\alpha} ~ \undersym{\lambda}{i}_{\beta}.
\end{equation}
It is easy to show that,
\begin{equation}\label{q5}
g^{\mu\alpha}g_{\alpha\nu} = \delta^{\mu}_{\nu},
\end{equation}
and
\begin{equation}\label{q6}
g \edf \|g_{\mu\nu}\| \neq 0 .
\end{equation}
The properties of $g_{\mu\nu}$ and its conjugate given by (\ref{q3}) - (\ref{q6}), show that this tensor can be used as a metric defining a Riemannian structure, associated with the AP-structure. \\
\paragraph{}
The AP-condition,
\begin{equation}\label{q7}
\undersym{\lambda}{i}_{\overset{\mu}{+}|\nu} = 0,
\end{equation}
implies the existence of a linear connection,
\begin{equation}\label{q8}
\Gamma^{\alpha}_{.\mu\nu} = \undersym{\lambda}{i}^{\alpha} ~ \undersym{\lambda}{i}_{\mu,\nu},
\end{equation}
which is the direct solution of (\ref{q7}). Here we use the strock ($|$) and the ($+$) sign to characterize tensors derivatives, using the connection (\ref{q8}), and the comma (,) is used for partial differentiation.
 \paragraph{}
 It is clear that the linear connection (\ref{q8}) is non-symmetric. Consequently, it has a torsion given by,
\begin{equation}\label{q9}
\Lambda^{\alpha}_{.\mu\nu} \edf  \Gamma^{\alpha}_{.\mu\nu} - \Gamma^{\alpha}_{.\nu\mu}.
\end{equation}
Now, since (\ref{q8}) is non-symmetric, so its dual $ \widetilde{\Gamma}{^{\alpha} _{.\mu\nu}} (\edf \Gamma^{\alpha}_{\nu\mu})$ and its symmetric part $\Gamma^{\alpha}_{(\mu\nu)} (\edf \frac{1}{2}(\Gamma^{\alpha}_{\mu\nu} + \Gamma^{\alpha}_{\nu\mu} )) $ will serve as linear connections, as well. Imposing a metricity condition on (\ref{q4}), we can define another symmetric linear conection, Christoffel symbol of the 2$^{nd}$ kind $\{{^{\alpha}_{\mu\nu}} \}$, as usually defined. Now, we can define the $3^{\texttt{rd}}$ order tensor,
\begin{equation}\label{q10}
\gamma^{\alpha}_{.\mu\nu} = \undersym{\lambda}{i}^{\alpha}~\undersym{\lambda}{i}_{\mu;\nu}
\end{equation}
where ($;$) is used to characterized covariant differentiation using Christoffel symbol. The tensor defined by (\ref{q10}) is the contorsion of the space. It is easy to derive the following relations (cf.\cite{Wanas2002}),
\begin{equation}\label{q11}
\Gamma^{\alpha}_{.\mu\nu} = \{{^{\alpha}_{\mu\nu}} \} + \gamma^{\alpha}_{.\mu\nu}.
\end{equation}
\begin{equation}\label{q12}
\Lambda^{\alpha}_{.\mu\nu} =  \gamma^{\alpha}_{.\mu\nu} - \gamma^{\alpha}_{.\nu\mu},
\end{equation}
\begin{equation}\label{q13}
\Lambda^{\alpha}_{.\mu\alpha} =  \gamma^{\alpha}_{.\mu\alpha} = C_{\mu}.
\end{equation}
The vector given by (\ref{q13}) is known as the basic vector of the space (cf. \cite{Mikhail1962}).
Using (\ref{q9}), (\ref{q10}), (\ref{q13}), a set of $2^{nd}$ order tensors has been defined \cite{Mikhail1962} .These tensors are necessary for any physical applications. The following table gives the definition of these tensors,
\newpage

\begin{center}
Table (1): Second Order World Tensors \cite{Mikhail1962}
\begin{tabular}{|c|c|}
  \hline
  &\\
  Skew-Symmetric Tensors & Symmetric Tensors \\ \hline
  &\\

 $ \xi_{\mu\nu} \edf {\gamma_{\mu\nu.}^{~~\alpha}}_{| \overset{\alpha}{+}}$ &   \\
 &\\
 $\zeta_{\mu\nu} \edf C_{\alpha}\gamma _{\mu\nu}^{~..  \alpha}$ &   \\ \hline
 &\\
 $\eta_{\mu\nu} \edf C_{\alpha}\Lambda ^{\alpha}_{.~\mu\nu}$ & $\phi_{\mu\nu} \edf C_{\alpha}\Delta^{\alpha}_{.~\mu\nu}$ \\
 &\\
  $\chi_{\mu\nu} \edf {\Lambda^{\alpha}_{.~\mu\nu}}_{|\overset{\alpha}{+}}$ & $\psi_{\mu\nu} \edf {\Delta^{\alpha}_{.~\mu\nu}}_{|\overset{\alpha}{+}}$ \\ &\\
  $\varepsilon_{\mu\nu} \edf {C_{\mu}}_{|\overset{\nu}{+}} - {C_{\nu}}_{|\overset{\mu}{+}}$ & $\theta_{\mu\nu}\edf {C_{\mu}}_{|\overset{\nu}{+}} + {C_{\nu}}_{|\overset{\mu}{+}}$  \\
  &\\
  $k_{\mu\nu} \edf \gamma^{\alpha}_{.~\mu\epsilon}\gamma^{\epsilon}_{.~\alpha\nu} - \gamma^{\alpha}_{.~\nu\epsilon}\gamma^{\epsilon}_{.~\alpha\mu}$ & $\varpi\edf \gamma^{\alpha}_{.~\mu\epsilon}\gamma^{\epsilon}_{.~\alpha\nu} + \gamma^{\alpha}_{.~\nu\epsilon}\gamma^{\epsilon}_{.~\alpha\mu}$
  \\ \hline
  &\\
    & $\omega_{\mu\nu} \edf \gamma^{\epsilon}_{\mu\alpha}\gamma^{\alpha}_{.\nu\epsilon}$ \\
    &\\
    & $\sigma _{\mu\nu} \edf \gamma^{\epsilon}_{.\alpha\mu}\gamma^{\alpha}_{.\epsilon\nu}$  \\
    &\\
    & $\alpha_{\mu\nu} \edf C_{\mu} C_{\nu}$ \\
    &\\
    & $R_{\mu\nu}   \edf  \frac{1}{2} (\psi_{\mu\nu} - \phi _{\mu\nu} - \theta_{\mu\nu}) - w_{\mu\nu}$ \\
    &\\
  \hline
\end{tabular}
\end{center}
In this table we use ${T^{\alpha}_{.~\beta\gamma}}_{|\overset{\delta}{+}}$ for ${T^{\overset{\alpha}{+}}_{.~ \overset{\beta}{+} \overset {\gamma}{+}}}_{{| \delta}}$ and $R_{\mu\nu}$ is Ricci tensor defined in terms of the tetrad $2^{nd}$ order tensors defined in Table (1).
\subsection[]
{AP-Space Structure for Cosmological Applications \footnote{In this article, we are going to use relativistic system of units $c = G = 1$, where c is the speed of light and G is Newton's gravitational constant.}}
\hspace{12pt}
Robertson \cite{Robertson1932} has derived the building blocks of two AP-structures with homogeneity and isotropy, i.e. satisfy the cosmological principle. Further investigation of the two structures \cite{Wanas1986} show that they have the types \cite{Mikhail1981} FOGIII and FOGI, respectively. In the present work we are going to use the 1$^{st}$ one whose structure is given by the matrix (coordinate system used, $x^{0} = t $, $ x^{1} = x$ , $x^{2} = y$, $x^{3} = z$ ),
\begin{equation}\label{q14}
   \undersym{\lambda}{j}^{\mu}  =
  \begin{array}{ll}
      \underset{\underset{\downarrow}{j(0,1,2,3)}}{{\mu}\rightarrow}& (0,1,2,3)   \\
      & \left(
                   \begin{array}{cccc}
       1 & 0 & 0 & 0 \\[0.3em]
       0 & \frac{i(1 - \frac{1}{4}kr^{2} + \frac{k}{2} x^{2})}{a} & \frac{i(\frac{k}{2}xy - k^{\frac{1}{2}}z)}{a} & \frac{i(\frac{k}{2}xz + k^{\frac{1}{2}}y)}{a} \\[0.3em]
       0 & \frac{i(\frac{k}{2}yx + k^{\frac{1}{2}}z)}{a} & \frac{i(1 - \frac{1}{4}kr^{2} + \frac{k}{2}y^{2})}{a} & \frac{i(\frac{k}{2}yz -k^{\frac{1}{2}}x)}{a} \\[0.3em]
           0 & \frac{i(\frac{k}{2}zx - k^{\frac{1}{2}}y)}{a} & \frac{i(\frac{k}{2}zy + k^{\frac{1}{2}}x)}{a} & \frac{i(1 - \frac{1}{4}kr^{2} + \frac{k}{2}z^{2})}{a} \\
                   \end{array}
                 \right),
     \\
  \end{array}
\end{equation}
where ($a$) is a function of time only,  $i \edf \sqrt{-1}$,  $h \edf \frac{1}{1+\frac{1}{4}kr^{2}}$,  $r = \sqrt{x^2 + y^2 + z^2}$ and  $k (= 0, +1, -1 )$ is the curvature constant. The covariant components of this tetrad are given in the matrix:
\\
\begin{equation}\label{q15}
  \undersym{\lambda}{j}_{\mu}  =
  \begin{array}{ll}
      \underset{\underset{\downarrow}{j(0,1,2,3)}}{{\mu}\rightarrow} & (0, 1, 2, 3)\\
      & \left(
                   \begin{array}{cccc}
       1 & 0 & 0 & 0 \\[0.3em]
       0 & -i(1 - \frac{1}{4}kr^{2} + \frac{k}{2} x^{2})a h^{2} &  -i(\frac{k}{2}xy - k^{\frac{1}{2}}z) a h^{2} &  -i(\frac{k}{2}xz + k^{\frac{1}{2}}y)a h^{2} \\[0.3em]
       0 & -i(\frac{k}{2}yx + k^{\frac{1}{2}}z)a h^{2} & -i(1 - \frac{1}{4}kr^{2} + \frac{k}{2}y^{2})a h^{2} & -i(\frac{k}{2}yz -k^{\frac{1}{2}}x)a h^{2} \\[0.3em]
           0 & -i(\frac{k}{2}zx - k^{\frac{1}{2}}y)a h^{2} & -i(\frac{k}{2}zy + k^{\frac{1}{2}}x)a h^{2} & -i(1 - \frac{1}{4}kr^{2} + \frac{k}{2}z^{2})a h^{2} \\
                   \end{array}
                 \right).
     \\
  \end{array}
\end{equation}
Using definitions (\ref{q3}), (\ref{q4}) we get the following non-vanishing components of the metric tensor of the Riemannian structure associated with the AP-structure(\ref{q14}),
\begin{equation}\label{q16}
\begin{split}
g^{00} &= 1,    g^{11} = g^{22} = g^{33} = -h^{-2} a^{-2}, \\
g_{00} &= 1,    g_{11} = g_{22} = g_{33} = -h^{2} a^{2}.
\end{split}
\end{equation}
Now, using the building blocks (\ref{q14}),(\ref{q15}) and definition (\ref{q8}) for the linear connection, we can get the following non-vanishing components of the tetrad tensor, using definition (\ref{q9})\cite{Wanas1989} \\
\begin{equation}\label{q17}
\begin{split}
\Lambda^{1}_{.10} &= \Lambda^{2}_{.20}  = \Lambda^{3}_{.30} = - \Lambda^{1}_{.01} = - \Lambda^{2}_{.02} = -\Lambda^{3}_{.03} = \frac{\dot{a}}{a}, \\
\Lambda^{1}_{.32} &= \Lambda^{2}_{.13} = \Lambda^{3}_{.21} = - \Lambda^{1}_{.23} = - \Lambda^{2}_{.31} = - \Lambda^{3}_{.12} = 2\sqrt{k} h,
\end{split}
\end{equation}
where the dot (.) represents differentiation w.r.t. time.
Now the only non-vanishing components of the basic vector $C_{\mu}$ (\ref{q13})is:
\begin{equation}\label{q18}
    C_{0} = -3 \frac{\dot{a}}{a},
\end{equation}
from which we can get a scalar $\mathcal{T}$ defined by,
\begin{equation}\label{q19}
    \mathcal{T} \edf \sqrt{g^{\mu\nu} C_{\mu}C_{\nu}},
\end{equation}
which, in the present case using (\ref{q16}),(\ref{q18}), gives,
\begin{equation}\label{q20}
    \mathcal{T} = 3\frac{\dot{a}}{a}.
\end{equation}
\subsection{GR in The AP-Space}
Many authors have attempted to solve problems of GR by constructing new theories. In most cases this procedure does not focus on the weakness in GR that causes the problem. In the present work we are going to deal with GR itself, written in the AP-geometry. This will enable us to focus on the weaknesses of the theory and find a solution, if any.
\paragraph{}
The field equation of GR can be written as:
\begin{equation}\label{q21}
    R_{\mu\nu} - \frac{1}{2}g_{\mu\nu}(R-2\Lambda) = -kT_{\mu\nu}
\end{equation}
where $R_{\mu\nu}$ is Ricci tensor, R is Ricci scalar, $\Lambda$ is the cosmological constant and $T_{\mu\nu}$ is the material - energy tensor. Ricci tensor and scalar can be evaluated, in context of the AP-geometry, following either: \\
(i) Using the definition,
\begin{equation}\label{q22}
 R_{\mu\nu}   \edf  \{^{~\alpha}_{\mu\alpha}\}_{,\nu} - \{^{~\alpha}_{\mu\nu}\}_{,\alpha} + \{~^{\epsilon} _{\mu\alpha}\} \{~^{\alpha} _{\epsilon\nu}\} - \{~^{\epsilon} _{\mu\nu}\} \{~^{\alpha} _{\epsilon\alpha}\}
\end{equation}
where $\{\}$ is Christoffel symbol constructed as stated above, by imposing a metricity condition  on (\ref{q4}). Or: \\
(ii) Using the definition \cite{Mikhail1962}.
\begin{equation}\label{q23}
 R_{\mu\nu}   \edf  \frac{1}{2} (\psi_{\mu\nu} - \phi _{\mu\nu} - \theta_{\mu\nu}) - w_{\mu\nu},
\end{equation}
where the second order tensors, on the R.H.S of this definition, are defined in Table(1).
However, (\ref{q22}) and (\ref{q23}) give identical results, since both are constructed from the building blocks of the AP-Space. Using (\ref{q22}) or (\ref{q23}) we can write the field equations of GR (\ref{q21}), in the context of AP-geometry.
\section{Effect of Torsion on The Dynamics of FRW-Model}
In the present Section we are going to investigate the effect of torsion on the cosmological solutions of Einstein field equation (\ref{q21}). For this purpose we use AP-structure (\ref{q14}). In this case we get the following set of differential equations corresponding to (\ref{q21}) with $T_{\mu\nu}$ is material-energy tensor of a perfect fluid (as usually done in standard cosmology cf.\cite{Garcia1999} ).
\begin{equation}\label{q24}
    3\left(\frac{\dot{a}}{a}\right)^{2} = 8\pi{\rho_{_0}} - \frac{3k}{a^{2}} + \Lambda ,
\end{equation}
\begin{equation}\label{q25}
    3\left(\frac{\ddot{a}}{a}\right) = -4\pi(\rho_{_0} + 3P_{0}) + \Lambda ,
\end{equation}
together with the conservation equation,

\begin{equation}\label{q26}
 \dot{\rho} + 3\frac{\dot{a}}{a}(\rho_{_0} +P_{0}) = 0 ,
\end{equation}
where ${\rho}_{_0}$ and $P_{0}$ are the proper density and the proper pressure of a perfect fluid, respectively, and the dot (.) is used to characterize differentiation with time, as stated above. Equation (\ref{q24}) and (\ref{q25}) are the same FRW-dynamical equations of standard cosmology. As we are going to investigate the sole effect of torsion on the dynamics, we take $\Lambda = 0$ since, as it is well known, $\Lambda$ can be used as an alternative to solve some of such problems. Also we are going to take $k = 0$ in order to facilitate comparison with standard cosmology. In this case (\ref{q24}),(\ref{q25}) will reduced to,
\begin{equation}\label{q27}
   \left(\frac{\dot{a}}{a}\right)^{2} =\frac{8}{3} \pi{\rho_{_0}},
\end{equation}
\begin{equation}\label{q28}
    \left(\frac{\ddot{a}}{a}\right) = -\frac{4}{3}\pi(\rho_{_0} + 3P_{0}),
\end{equation}
which are the dynamical equations, of FRW standard cosmology, written in the context of the AP-geometry ( a geometry with non vanishing torsion ). It is well known that all solution of (\ref{q27}) and (\ref{q28}) have problems e.g. singularity, horizons, ...etc. We are going to insert the torsion  of the AP-space, to study its effect on the dynamics of the resulting World models. This can be done by considering the scalar torsion defined by (\ref{q19}) . For the AP-structure (\ref{q14}), it has the following non-vanishing value,
 \begin{equation}\label{q29}
   \mathcal{T} = 3 \frac{\dot{a}}{a},
\end{equation}
which can be related to Hubble's parameter (H)as:
\begin{equation}\label{q30}
\mathcal{T} = 3H.
\end{equation}
Now, we have,
 \begin{equation}\label{q31}
   \dot{\mathcal{T}} = 3\dot{H} = 3\left[\frac{\ddot{a}}{a} - (\frac{\dot{a}}{a })^{2} \right],
\end{equation}
and then,
\begin{equation} \label{q32}
\frac{\ddot{a}}{a} = \frac{1}{9}\mathcal{T}^{2} + \frac{1}{3}\dot{\mathcal{T}} .
\end{equation}
This relation gives the effect of the scalar torsion on the acceleration. Now comparing (\ref{q27})with (\ref{q30}) we get,
\begin{equation}\label{q33}
\rho_{_ {_{_\mathcal{T}}}} = \frac{1}{24 \pi}\mathcal{T}^{2}.
\end{equation}
Again, comparing (\ref{q28})with (\ref{q32}) and using(\ref{q33}) we get,
\begin{equation}\label{q34}
P_{_ {_{_\mathcal{T}}}} =  - \rho_{_ {_{_\mathcal{T}}}}[ 1 + \epsilon ].
\end{equation}
where $\rho_{_ {_{_\mathcal{T}}}}$ , $P_{_ {_{_\mathcal{T}}}}$ are the proper density and the proper pressure corresponding to the scalar torsion field $\mathcal{T}$,  and
\begin{equation}\label{q35}
\epsilon \edf  \frac{2\dot{\mathcal{T}}}{\mathcal{T}^2}.
\end{equation}
\textbf{From equation (\ref{q34}), it is clear that we have the following possibilities:}\\
(i) when $ \epsilon  = 0 $, giving $P_{_ {_{_\mathcal{T}}}} = - \rho_{_ {_{_\mathcal{T}}}}$ and $w(\edf\frac{P}{\rho}) = -1$, which means that $ \epsilon \edf  \frac{2\dot{\mathcal{T}}}{\mathcal{T}^2} = 0 $, then $\dot{\mathcal{T}} = 0 $.  So,  from equation (\ref{q30}), we get:
\begin{equation}\label{q36}
 H = \frac{\dot{a}}{a} =  \frac{1}{3} \mathcal{T}  =  const.
\end{equation}
This case corresponds to the solution,
\begin{equation}\label{q37}
a = e^{\frac{1}{3}\mathcal{T} t},
\end{equation}
which corresponds to an exponentially expanding Universe ( a solution required for inflation cf.\cite{Guth1981, Linde1982, Guth2005}, dark energy cf.\cite{Li2012}).\\
(ii) $\epsilon>0$ this also gives negative pressure with $w < -1$, which leads to the phantom accelerating universe "Super-Negative Equation of State" (cf.\cite{Caldwell1999, Caldwell2003, Sharif2011}). A phantom energy component, as "Robert etal.\cite{Caldwell2003} stated:\emph{"something which is apparent to the sight or other senses but has no corporeal existence – an appropriate description for a form of energy necessarily described by unorthodox physics, as will be seen later"} .

(iii) $ \epsilon  < 0 $, and $ 0 > w > -1 $, the universe in quintessence era \cite{CALDWELL2000, Bassett2007, Hossain2012, Ali2009}. The difference between the quintessence model of Dark Energy and the cosmological constant is that the quintessence can vary with time \cite{Bassett2007},


\section{Discussion and concluding Remarks:}
\hspace{12pt}
The present article represents an attempt to diagnose the weakness of GR, which leads to famous problems of standard cosmology. We begin the discussion by clarifying an important point.   \\
In the present work we used the AP-geometry to write the field equation of GR. In general, as stated in the text, the structure of an AP-space in 4-dimensions, is defined completely by 16-independent building blocks (functions). So, one needs 16-conditions (equations) to find these function. GR field equations (\ref{q21}) represent 10 of such conditions. Then we need 6 more conditions to fix the functions. Fortunately, in the AP-structure with homogeneity and isotropy (\ref{q14}), all skew $2^{nd}$ order tensors, of Table (1) vanish identically. And since the 6-conditions needed are expected to be given by a tensor equation including some combinations of the skew tensor of Table (1), then we can say that the required six conditions are satisfied identically for the AP-structure(\ref{q14}).
\paragraph{}
GR is a geometric theory for the gravitational interaction. It gained most of its success from the applications of its field equations (in \textbf{free} space) together with the geodesic equations. Such successful applications are carried out in the cases of spherical symmetry (The Schwarzschild solution) and axial symmetry (Kerr solution). In this sense, GR, constructed in Riemannian geometry, can be considered as a \textbf{pure} geometric theory for gravity, since all entities of the theory are built from the building blocks of the geometry (the metric tensor).
\paragraph{}
In the contrast, many problems appear when using the field equation within a \textbf{material distribution}(\ref{q21}), for applications. The main difference between GR in free space and in a material distribution is the tensor $T_{\mu\nu}$, which has no representative in Riemannian Geometry. In this sense, GR, within a material distribution, cannot be considered as a pure geometric theory for gravity.
\paragraph{}
It is to be noted that Riemannian geometry has no sufficient structure to accommodate any physical entities but the gravitational field and its background of space-time. So, on one hand, one has to use a more wider geometry than the Riemannian one, in order to represent more physical quantities, e.g. matter and energy, and to write a pure geometric theory. On the other hand, a more wider geometry would have torsion in addition to curvature as the two important geometric object characterizing the geometry. In this case what would be the impact of torsion on the solutions and predictions of any suggested field theory, including GR, written in the context of this geometry?
\paragraph{}
The present work, gives an answer to the above question. Some of the problems of standard cosmology have been solved (disappeared) e.g. the solution (\ref{q37}) has no singularity and no particle horizon problems. The scheme leading to this achievement can be summarized as follows. The field equations of GR (\ref{q21}) are written in the AP-geometry and applied to an AP-structure having homogeneity and isotropy (\ref{q14}). The resulting differential equations (\ref{q24}),(\ref{q25}) are FRW-dynamical equations, characterizing standard cosmology. Till this step, nothing is changed concerning standard cosmology and its problems still exist. Inserting the effect of the AP-torsion associated with AP-structure (\ref{q14}),  we get the following results:\\
1 - we arrived to the possibility of (-ve) pressure leading to,

 ~ ~ ~ ~ ~ - inflation

 ~ ~ ~ ~ ~ - accelerating expansion of the Universe with

 ~ ~ ~ ~ ~ ~ ~ ~ ~ ~ ~ ~ ~ ~ ~ ~ ~ ~   - phantom energy

 ~ ~ ~ ~ ~ ~ ~ ~ ~ ~ ~ ~ ~ ~ ~ ~ ~ ~   - quintessence.\\
2 - Torsion field is closely related to Hubble's parameter, equation (\ref{q30}).\\
3 - The results similar to inflation are obtained without any need to any of the following assumptions:

   ~ ~ ~ ~ ~ - A Lagrangian density .

   ~ ~ ~ ~ ~ - The Potential Energy term dominates over the Kinetic Energy one.

   ~ ~ ~ ~ ~ - Minimal coupling.

   ~ ~ ~ ~ ~ - The existence of any object from outside the geometric structure.

   ~ ~ ~ ~ ~ - Imposing homogeneity and isotropy on the torsion field.\\
4 - GR can be considered as a pure geometric theory since all geometric objects, including $T_{\mu\nu}$, are now constructed from the building blocks of the AP-structure. More specifically, the material- energy tensor in the present work corresponds to the torsion field. The matter is not an ordinary one in this case.\\
5 - In the case of $P_{_ {_{_\mathcal{T}}}} = - \rho_{_ {_{_\mathcal{T}}}}$,  i.e. $w = -1$, if we substitute into the conservation equation (\ref{q26}) we get,
   \begin{equation}\label{q38}
   \dot{\rho_{_ {_{_\mathcal{T}}}}} = 0
   \end{equation}
   This result is somewhat puzzling. Although we are using the conservation equation (\ref{q26}) we get a constant density in an expanding universe!,  and consequently a constant pressure!. It seems that torsion field, which is absent from the conservation equation, is converted into a type of matter and energy. To understand this process, torsion energy \cite{wanas2007} should be include in the conservation equation (Conservation law should be modified in order to include torsion energy). This work is in progress now.

\end{document}